# Designing an Android Application for Bills Segregation

Ruifeng Guo

ma2808203259@hotmail.com

https://github.com/ma2808203259/SplitBill

## Abstract
In recent years, several people have been hanging out or sharing rental house with others. For utilities bills and other items, most people have been using the tedious way where a person pay the bill and others transfer the money to their bank account. We research and develop android app as there has been a few companies trying to solve this. Our android application for bills segregation can resolve this problem. In our novel system, the user can create "event" which represent all events that required to split the payment to everyone. In this application, the user can invite the people and set the rules such as the percentage of each person. More importantly, the users will have the dashboard to display all the event and be able to chat with each other seamlessly.

## Introduction
The document proposed an android application called 'Separate Bills' for people to split their bills such as electric power, water and lunch. The language for developing this application is Kotlin [1]. This document will present the background and summary of this application, introduce the market research and what is my motivation to build this, point out the competitor which already exists in the market. Then this document would deep into the features of the application, list all the assets that would be used to finish this project, and very important analysis my target audience and the UI/UX design with coding concept explains. This document then will analyse features based on the user stories. Finally, will summary of the whole large system and schedule the sprints.

### Background
We are now living in a community, every week we may have several times to hang out or having dinner with friends. However, usually when we pay the fees for dinner or events. We make one person pay for that, and later everyone transfers their part into that person's bank account. Also, in another situation, not everyone can afford a house to live in, specifically the young generation. So, many people choose to rent a house with others. Therefore, the way they pay the bills such as water, gas and power would be the same as the last situation. This sometimes could be inconvenient.

### Overview
So, my application is designed to overcome this kind of problem. 'Separate bills' can move that action into the digital world to make it more efficient and convenient. It achieves that by creating events with the description of the event (is a dinner check or water bill?), total cost, people that have to pay this event and the 'rules' for this event which is the proportion for everyone. The user can add friends and chat directly in the application and the invitation will be received via message. After the user confirmed an event, it will show on the homepage until the user paid. The user can link multiple cards to pay the bills.

## Market research and motivation

During my research, the restaurant can split the bills for you manually or using technology. Some of the restaurants can provide to manually split bills for customers and these days more and more restaurants are bringing in the technology solution to help them with that. For example, the 'SplitAbility Pos' team provide an IT solution to help restaurants deal with that problem [1]. For the home bills, the payment giant companies such as PayPal, Alipay have their services for split the home bills such as water and gas. Their logic is simple evenly split the amount base among the number of people.

The motivation for me to develop this application because as an international student, I am renting a house and share with my friends. So, I faced to split bill situation every month. And, I like to hang out with my friends. We usually had dinner 2~3 times per month before the Covid-19. Every time we split the dinner bills by the traditional way, transfer money to the payer's bank account. This drives me to find out a solution to solve this problem.

## Competitor analysis

As I mentioned in the market research part, the restaurant and the payment service companies have their solution to overcome in their area. However, their solutions all have limitations. First, on the restaurant side, not all restaurants would update their checking system to compatible with the split bill function, because it costs and not that necessary. Even the restaurants who got this system somethings they are not willing to use, because something called 'time cost'. Thos Weatherby mentioned that in Quora. [2] On the other hand, the payment companies' function to split the home bills. For example, PayPal's split bill function is inserting the amount and the people them it will automatically evenly generate the cost for each person. I will say it works well and only works for home bills. Also, I believe everyone how is using PayPal must know its fees, painful. So, I believe only the people who use PayPal normally will use that function.

My application compares to them has a huge advantage. My application treats everything that needs a split bill as an event. The user can make the rules like the proportion for everyone. And it does not require the business to pay thousands of dollars for a split bill system.

## Features

### Asset list

The tools would be used in this project:

- Figma: Figma is a vector graphics editor and primarily web-based prototyping tool, so it is very simple to use and very easy to share the prototype with team members or showcase to the customers.
- Font-generator.com: font-generator is a website to generate different font styles for the logo. It is a free website, but it will have a watermark on the image.

- Android Studio: Android Studio is the official integrated development environment for Google's Android operating system. This is an indispensable tool to develop an android application.
- MongoDB: MongoDB is a source-available cross-platform document-oriented database program. MongoDB is also a NoSQL database program. It is free and has cloud containers.
- Kotlin: Kotlin is a modern mobile app programming language that is one of android studio accepts. Compare to Java for beginners, to achieve the same function, Kotlin has much fewer lines to write. Therefore, it is easy to write and debug for beginners.
- Trello: Trello is a web-based, Kanban-style, list-making application. Trello is very useful for project management. I can set few sections such as ToDo, Done and Planned. Setting the estimated start and finish date to manage the project schedule.
- GitHub: GitHub is a website that provides cloud storage and version control for our project via Git. I will use GitHub from the beginning of my project. It helps me to manage my project.
- Android studio emulator: android studio emulator will be used to debug the code in android studio. We can have multiple versions of android and multiple phones emulators.
- Microsoft teams: Teams is a chat software; it will use to communicate with unit staffs for getting help.
- Real android device: android studio provides the function to test the code on real android devices. So, after all works have done it will be used to make sure it can run on real devices.
- Flutter: Flutter is an open-source development kit. I will use flutter to make my interface pretty.
- Visual studio code: VS code is a very friendly coding editor, sometimes I will use it to write code.

## Product purpose

### Target audience

I think it is obvious in the project introduction. My application's target audience is the young generation who have not got their own house yet and like to hang out with friends. By using this application, they can do all split fees together including lunch or dinner fees, home bills and any event that must share the price such as tickets to watch football, you name it. Also, they can save time for both themselves and the business such as the restaurant, and they do not have to pay PayPal the high transfer fees. Even more, they can chat with friends in this application, or remainder someone who has not to pay the bill.

### Unique design and functions

My application combined the daily entertainment fees and home bills. There were many solutions for each one of them, however, I think their solutions are expensive and not efficient. Also, I think inside a payment application chat function might be convenient. Imagine if you transfer money to your friend via PayPal and then open up WhatsApp to tell them. Is that a bit complex?

### Explanation of features

Figma URL: https://www.figma.com/proto/AguYGhvPIGCntcAij4JiQw/separate-bills?node-id=2%3A28&scaling=scale-down&page-id=0%3A1

## User stories

| Title | Acceptance criteria | Priority | Estimate |
|---|---|---|---|
| **As a user, I want to signup if first using, login if I have an account so I can start to use the application.** | Give two options when open the application: login and signup.<br>When the user tap on the login then take him/her to the login page<br>When the user tap on the signup take him/her to the signup page | Medium | 8 |
| **As a user, I want to see my event on the homepage, so I can know which event I have now and be able to pay.** | Give the current events display on the user's homepage.<br>When the user clicks on any event shows the detail and pay button of that event.<br>When the user tap on the payment button, take him/her to the payment page and then checkout. | High | 8 |
| **As a user, I want to host an event so I can request my friend to pay me.** | Give a button for create host at all the time.<br>When the user tap on that, then lead the user to create an event.<br>And then send notifications for all users that invited via chat. | High | 8 |
| **As a user, I want to have a place to chat so I can receive invitations and communicate with my friend.** | Give a button at the bottom bar to navigate the user to chat.<br>Display all chat history for the user.<br>When the user clicks on one chat display the chat window.<br>User can receive notifications via chat | Medium | 8 |
| **As a user, I want to set my profile, so I can get personalise profile and manga my credit cards** | Give a button to the user profile page.<br>Give the setting button.<br>When the user tap on the setting button led the user to the setting page.<br>Allow the user to edit his profile and manage cards on this page. | High | 8 |
| **As a user, I want to search and add friends so I can invite them to an event.** | Give a button to the search page.<br>Allow user to search by username.<br>When the user finds the right person gives an option to send add friend request. | Medium | 5 |

## UI/UX design and components

*login or signup*

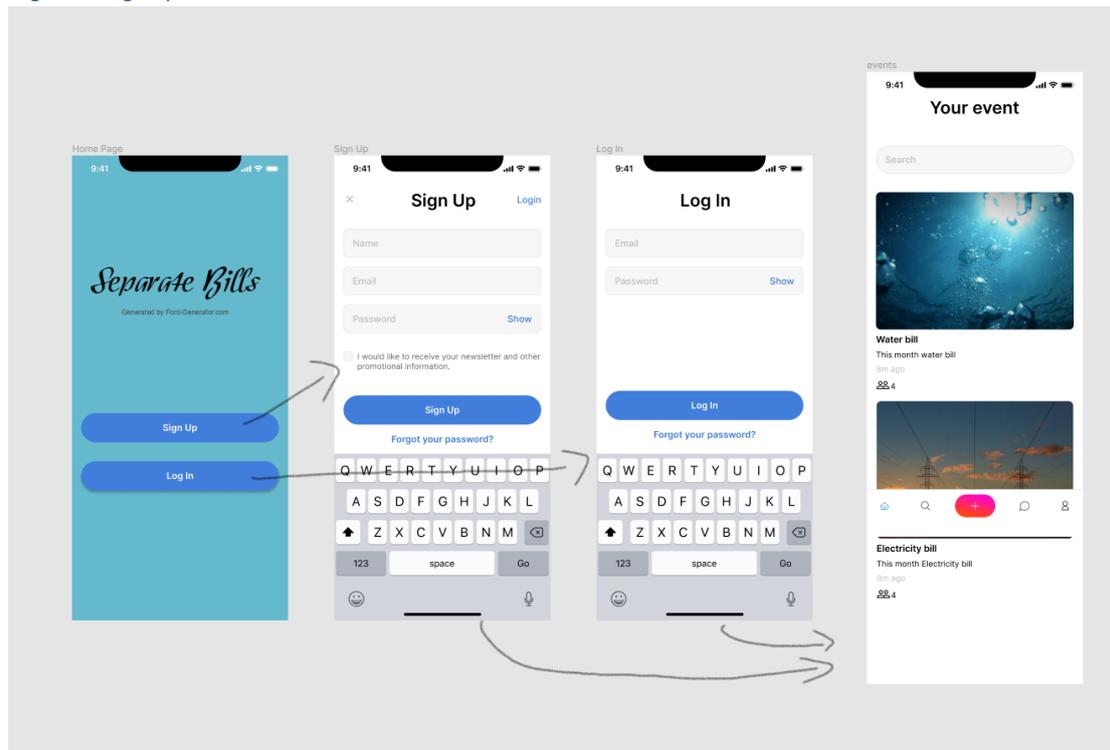

*Figure 1 Login or signup components*

For login or signup, I have designed 4 pages: start page, signup page login page and homepage.

Start page: this page for user first time uses this application.

Signup page: For people to sign up here if they have no account.

Login page: For people who already have an account.

Homepage: This page is to display the current event that the user. Will explain more in the next section.

Component explanation: This component is to make the user create an account or login using an existing account. The start page has a logo, and two buttons redirect to the signup page and login page. By tapping the signup button user can create an account by providing a name, email and password. After signup success, the user would redirect to the homepage. By clicking the login button user can log in by email and password. When the login success, the homepage would display.

*Pay the event.*

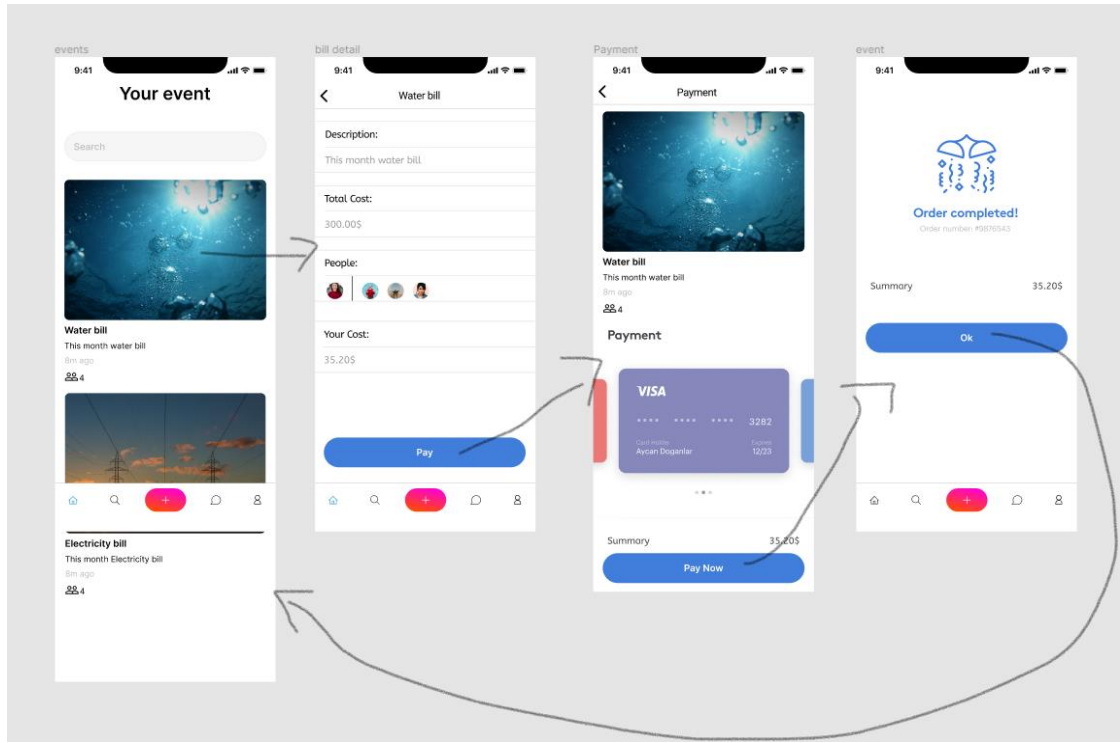

*Figure 2 Pay event components.*

Homepage: This is the homepage display all the events that need to pay

Event detail page: This is the event detail page, which shows details for the event.

Payment page: Display payment details.

Result page: Tell user their payment done.

Component explanation: This component is to check the current events and make a payment. On the homepage, the user can go through all the current events. By tapping one of them, the detail shows to the user, there has the description of this event, total cost of this event, members of this event and how much should use to pay. Also, please noticed in the people section, one people are separated with the other three which means she/he is the event host. Then the payment page, the user can see the event, amount and choose a card to pay. After the payment success, the system will have a result page to remind the user.

*Host event*

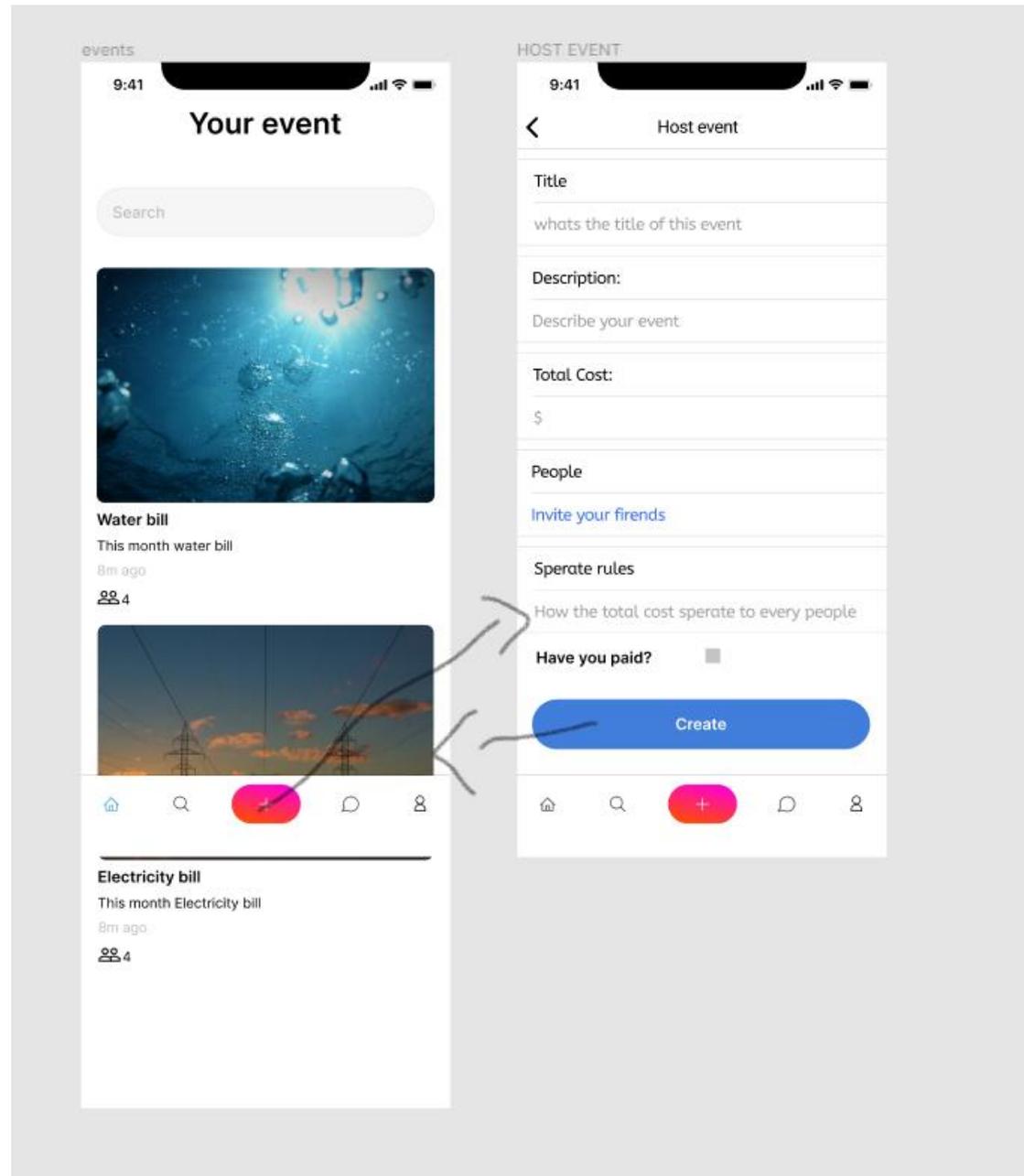

*Figure 3Host event component*

Homepage: This is the homepage display all the events that need to pay

Create event page: this page is used for creating a new event.

Component explanation: By clicking the red plus button this user will navigate to the host event page. It required the user to fill in the title of the event, the description, the total cost, the people who need to pay and the rules for this event (proportion for everyone). After creating the event the user would be taken back to the home page and the notification would be sent.

*Chat and add friends.*

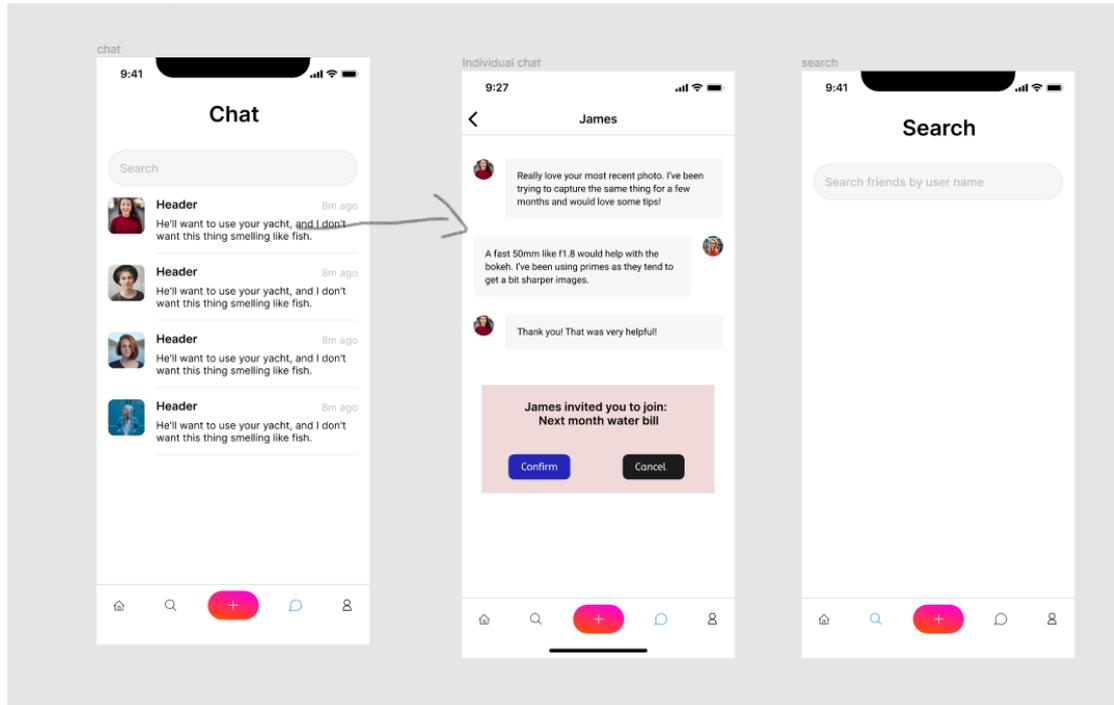

*Figure 4 chat and add friend.*

Chat page: Show all chats.

Individual chat page: Show chat window.

Search page: search for friends.

Component explanation: The chat page displays all user's chat, and clicking one of them will display the logs for that person. The chatbox also used to receive the invitation, as you can see the user can confirm which will add an event on the homepage or cancel it. The search page is used to search and add new friends by username.

*User profile and settings.*

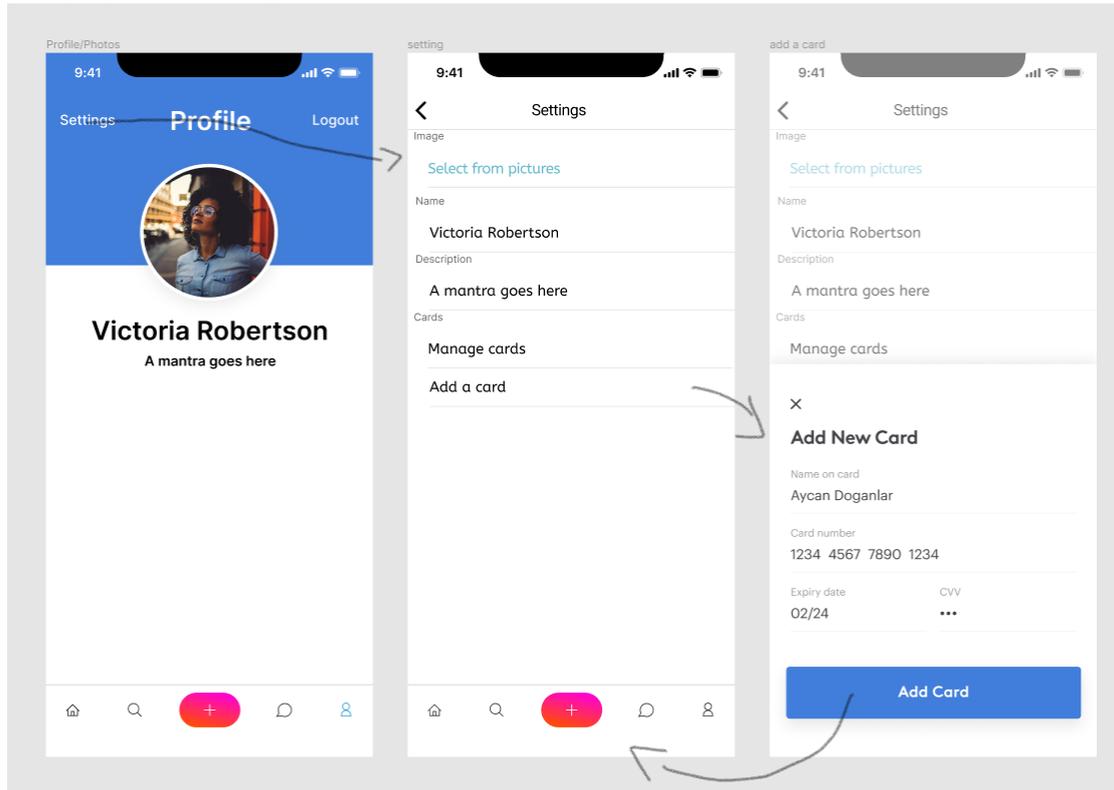

*Figure 5 profile and settings*

Profile page: To display the user profile such as icon name.

Setting page: edit the user profile and manage the cards.

Add card page: Add a card for payment.

Component explanation: The profile page can display the information of the logged user, the user can logout here or go to settings. On the setting page, the user can edit his/her profile such as name. Also, the user can manage cards or by tapping add a card to the add card page. Fill in the card details and link a new card.

Larger System overview

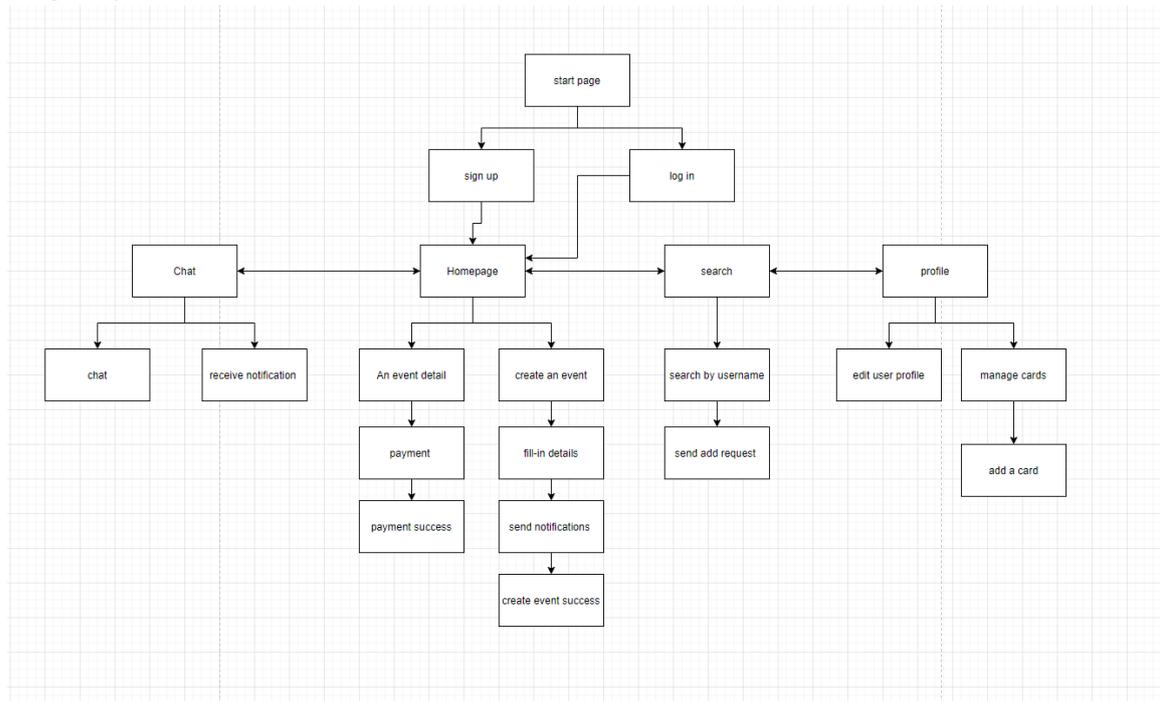

*Figure 6 workflow*

As you can see in the figure, it is the workflow of the application. There always a bottom bar for chat, homepage, search and profile. So, these four nodes in figure6 are in the same layer and connected to each other. The chat node can do chat and receive notifications, the homepage can make a payment or create an event, the search can search and add friends and the profile node can edit profile or manage the cards.

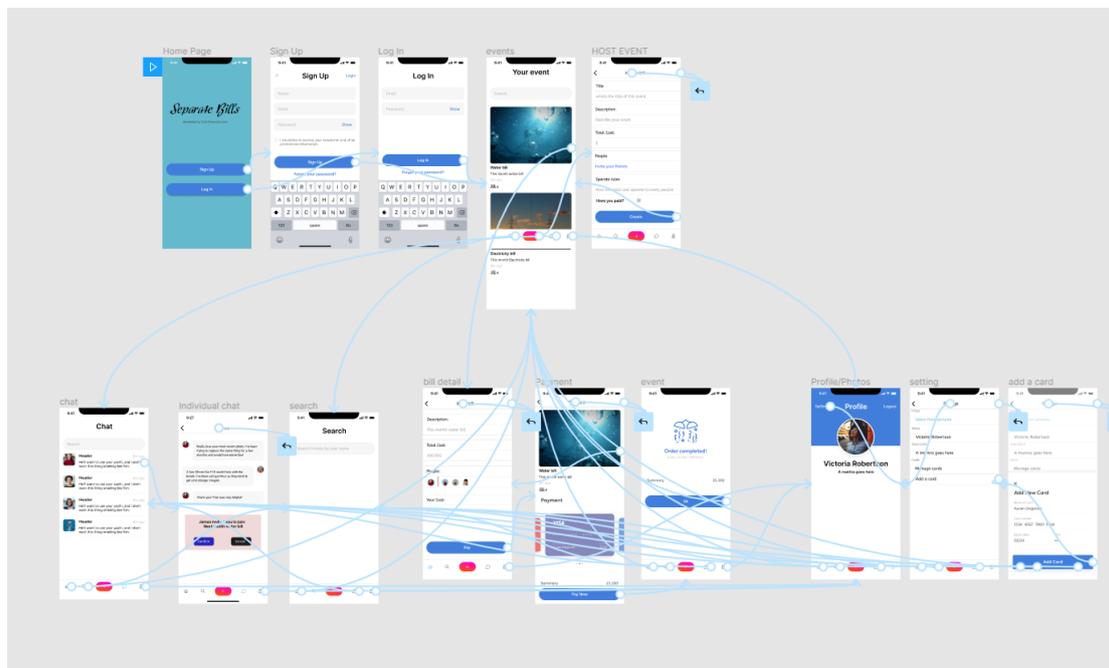

*Figure 7 wireframing*

### Coding concepts

- Classes and Object: A class is a blueprint for creating objects. It has member functions and member variables. and an object is an instance of the class. One can access class member functions and variables with the help of an object. [3]
- OOPS (Object Oriented Design): It is a methodology or paradigm to design a program using classes and objects. The Main Components of OOPS are Encapsulation, Abstraction, Inheritance and Polymorphism [3]
- Navigation: Navigation is the act of moving between screens of an app to complete tasks.
- Fragments: A Fragment represents a reusable portion of your app's UI. A fragment defines and manages its layout, has its lifecycle, and can handle its input events. [4]
- NoSQL: NoSQL databases are purpose-built for specific data models and have flexible schemas for building modern applications. [5]
- Collections: Kotlin collection library provide such as list set and map, help to dual with data.
- Library: Collection of precompiled routines that we can use.

## Project plan

I split the project plan into 3 sprints:

Sprint1:

Week1 ~4

This sprint is to start the project, research online gets ideas and prepare the project proposal, using Figma to build the prototype for the showcase.

Sprint2:

Week5~8

This sprint aims to almost finish all the XML files which is the design and start to write backend functions. Playing around with database etc.

Sprint3:

Week9~12

This sprint aims to finish the project, finish all the functions and test the application on both emulator and real device. Finish the handover document and present the application.

Performance Analysis and Future Directions

We will perform rigorous testing of our application over transport layer protocols including TCP [6] [7] [8]and multipath TCP [9] [10], among others over widely used WiFi [6] [7] and Cellular networks [9]. For future enhancements we will seek ways to plug and play our apps for advances in distributed computing such as federated learning [7] and distributed ledger [11].